\documentclass[conference,a4paper,10pt]{IEEEtran}
\IEEEoverridecommandlockouts
\usepackage{cite}
\usepackage{flushend}
\usepackage{amsmath,amssymb,amsfonts}
\usepackage{algorithmic}
\usepackage{graphicx}
\graphicspath{ {./pngs/} }
\usepackage{textcomp}
\usepackage{xcolor}
\usepackage{pbalance}
\def\BibTeX{{\rm B\kern-.05em{\sc i\kern-.025em b}\kern-.08em
    T\kern-.1667em\lower.7ex\hbox{E}\kern-.125emX}}
\let\oldtt\texttt
\renewcommand{\texttt}[1]{\small\oldtt{#1}}

\renewcommand{\verb}[1]{\small\oldverb#1}

\newlength{\footnoterulewidth} 
\newlength{\footnoteruleheight} 
\renewcommand{\footnoterule}{   
    \kern -3pt   \hrule width \footnoterulewidth height \footnoteruleheight   \kern \dimexpr 3pt - \footnoteruleheight \relax 
}

\begin{document}

\title{
    Matrix Profile based Anomaly Detection in Streaming Gait Data for Fall Prevention
\thanks{This work has been supported by Estonian Research Council, grant No PRG424.}
}

\author{
    \IEEEauthorblockN{
        Branislav Gerazov$^1$, Elena Hadzieva$^2$, Andrei Krivošei$^3$, Fiorella Ines Soto Sanchez$^3$, \\
        Jakob Rostovski$^3$, Alar Kuusik$^3$, and Mahtab Alam$^3$
    }

    \IEEEauthorblockA{
        \\
        $^1$  FEEIT, Ss Cyril and Methodius University, Skopje, N Macedonia\\
        $^2$ University of Information Science and Technology ``St. Paul the Apostle'', Ohrid, N Macedonia\\
        $^3$ Tallinn University of Technology
      \\
        \\
        \texttt{gerazov@feit.ukim.edu.mk, elena.hadzieva@uist.edu.mk, alar.kuusik@taltech.ee}
    }
}

\maketitle

\begin{abstract}
The automatic detection of gait anomalies can lead to systems that can be used for fall detection and prevention.
In this paper, we present a gait anomaly detection system based on the Matrix Profile (MP) algorithm. 
The MP algorithm is exact, parameter free, simple and efficient, making it a perfect candidate for on the edge deployment.
We propose a gait anomaly detection system that is able to adapt to an individual's gait pattern and successfully detect anomalous steps with short latency.
To evaluate the system we record a small database of enacted anomalous steps.
The results show the system outperforms a more complex Neural Network baseline.
\end{abstract}

\begin{IEEEkeywords}
gait, anomaly, matrix profile, fall detection, edge
\end{IEEEkeywords}

\section{Introduction}
Certain neurological disorders reflect on an individual's ability to maintain stable gait.
This can lead to falls and cause significant physical, emotional and financial setbacks for the individual and their family, as well as a burden to health-care providers \cite{2011}, \cite{gait-detect}. 
Even though 46$\%$ of neurological patients fall at least once a year, potential predictors of falls are poorly investigated and understood \cite{ehrhardt}. 

There are two general approaches to analyzing the causes of falls: fall risk assessment through clinical investigations \cite{Schniepp,  ehrhardt, Leddy, Vance}, and computerized gait analysis \cite{2011, gait-detect, Meyer, Monoli, Gao}. 
In \cite{ehrhardt}, the authors distinguish fallers from non-fallers among neurological patients, based on spatio-temporal, variability and asymmetry gait parameters. 
Similarly, \cite{Meyer} make a retrospective classification between fallers and non-fallers among patients with Multiple Sclerosis based on accelerometer and gyroscope data, applying deep learning models. 
The desire is to develop early, automatic prediction of missteps that might cause falling and a way to intervene and prevent it.

The wrong step in one's gait is an anomaly, or outlier, in the sequence of normal steps \cite{Grubbs,Ahmad,Munir}. 
Detecting anomalies in streaming data is a challenging task: $(i)$ the stream is infinite, which makes storing the entire stream impossible; $(ii)$ the stream contains mostly normal instances and much less anomalies; and $(iii)$ streaming data evolves over time, imposing the need for adaptation \cite{Tan, Ahmad}. 
When dealing with anomalies in gait, there's an additional challenge in that there is both interpersonal variability, i.e. each person's gait is unique, as well as intrapersonal variability as one's gait is not set in stone.

Solving the problem requires a robust algorithm that will work on streaming data, in an unsupervised and automated fashion, and that will be able to detect the anomaly with the highest possible accuracy as early as possible, a problem termed early classification of time series \cite{teaser}. 
Many anomaly detection algorithms exists, supervised  and unsupervised, yet the vast majority of them are unsuitable for real-time streaming applications \cite{neurocomp}.  
Moreover, algorithms operating on small data, e.g. shapelets \cite{gupta}, are still in its nascence.

In this paper, we present a gait anomaly detection system based on the Matrix Profile (MP) algorithm \cite{MPI}. 
The MP algorithm is exact, simple and parameter free, with low complexity.
Additionally, it is shaplet-based and thus interpretable \cite{Ji}. 
We first explore the plausibility of using the MP as a basis for a gait anomaly detection system and then develop it's design.
To evaluate the system's performance we record a small database of enacted anomalous steps.
Finally, we compare the proposed system to a more complex Neural Network baseline.

\section{Matrix Profile}

The following definitions of the MP are slightly modified from \cite{MPI, MAD} and \cite{MERLIN}, in favor of mathematical correctness and conciseness. 
A {\bf time series} $ \mathbf{T}=\{t_k\}_{k=1}^n$ is a sequence of $n$ real values. 
The sub-sequence of $m$ consecutive terms of $\mathbf{T}$, starting from the position $i$, where $1\leq i \leq n-m+1$, will be denoted by $\mathbf{T}_{i, m}$. 
Thus, $\mathbf{T}_{i, m}=\{t_k\}_{k=i}^{i+m}$. 
The sub-sequences will be compared using the {\it $z$-normalized Eucledian distance}. 
An {\bf all-subsequences set $\mathbf{A}_{\mathbf{T}}$} of a time series $\mathbf{T}$ is an ordered set of all possible sub-sequences of $\mathbf{T}$ obtained by sliding a window of length $m$ across $\mathbf{T}$: $\mathbf{A}_{\mathbf{T}} = \{\mathbf{T}_{1, m}, \mathbf{T}_{2, m}, \ldots, \mathbf{T}_{n-m+1, m}\}$. 
The {\bf matrix profile} (MP) is a vector of length $n-m+1$ corresponding to all-sub-sequences set, whose $i$-th location is the distance of the sub-sequence $\mathbf{T}_{i, m}$, to its nearest neighbor, under $z$-normalized Euclidean Distance. 

Note that trivial matches are avoided, that is the sub-sequences that overlap at least in the half length with $\mathbf{T}_{i, m}$ are not taken into account in computing the $i$-th component of the matrix profile of $\mathbf{T}$ (\cite{MPI}). 
Given a time series $\mathbf{T}$, the sub-sequence $\mathbf{T}_{i, m}$ is said to be the {\bf discord} of $\mathbf{T}$ if $\mathbf{T}_{i, m}$ has the largest distance to its nearest (non-trivial) match.

In the whole algorithm there is only one parameter to set -- the length of the sub-sequence. 
In our application scenario, this would correspond to the length of a single step.
Extracting the motif, i.e. reoccurring pattern, and discord from our generated gait data set means extracting the normal and anomalous step, correspondingly. 
We used the STAMP (Scalable Time series Anytime Matrix Profile) \cite{MPI} and STOMP (Scalable Time series Ordered-search Matrix Profile) \cite{MPI, STOMP} algorithms for generating the matrix profile and detecting both motif and discord of particular time series.  

\section{Dataset}


We recorded a small dataset that includes anomalous steps dispersed amidst normal walking patterns by a single male subject on a hard surface.
The anomalous steps were meant to mimic pathological step patterns from different disorders. 
In total 9 pathological step patterns were included in the dataset as shown in Table~\ref{tab:dataset}.
The recording protocol comprised of walking a straight line of around 10 steps and acting out a pathological step pattern in the middle.
Each recording contains around 4 stretches of 10 steps. 

We recorded the data using a Shimmer Inertial Measurement Unit (IMU) sensor placed on the foot of the subject that records accelerometer and gyroscope signals in the 3 axes \cite{Jarchi}.
All the data was annotated in a two-step process:
$(i)$ steps were automatically segmented, q.v. Sec \ref{sec:step}, and
$(ii)$ the segments were manually corrected and labeled with three labels: ``ok'' for a normal step, ``ab'' for an anomalous step.

\begin{table}[t]
    \caption{Dataset of anomalous steps recorded for the analysis.}
    \label{tab:dataset}
    \begin{center}
    \begin{tabular}{lcccc}
        Pathology & recordings & ok & ab & duration [min] \\
        \hline\hline
        Antalgic & 7 & 156 & 42 & 6.61 \\
        Ataxic & 4 & 82 & 27 & 3.90 \\
        Diplegic & 6 & 139 & 36 & 6.05 \\
        Hemiplegic & 5 & 100 & 28 & 6.61 \\
        Hyperkinetic & 4 & 95 & 30 & 4.04 \\
        Parkinsonian & 4 & 100 & 30 & 4.07 \\
        Slap & 5 & 105 & 37 & 4.59 \\
        Steppage & 9 & 185 & 58 & 7.05 \\
        Trendelenburg & 4 & 85 & 30 & 4.13 \\
        \hline
        Total & 48 & 1047 & 318 & 44.60 
    \end{tabular}
    \end{center}
\end{table}

\section{Plausibility}

\begin{figure}[t]
    \includegraphics[width=0.9\columnwidth]{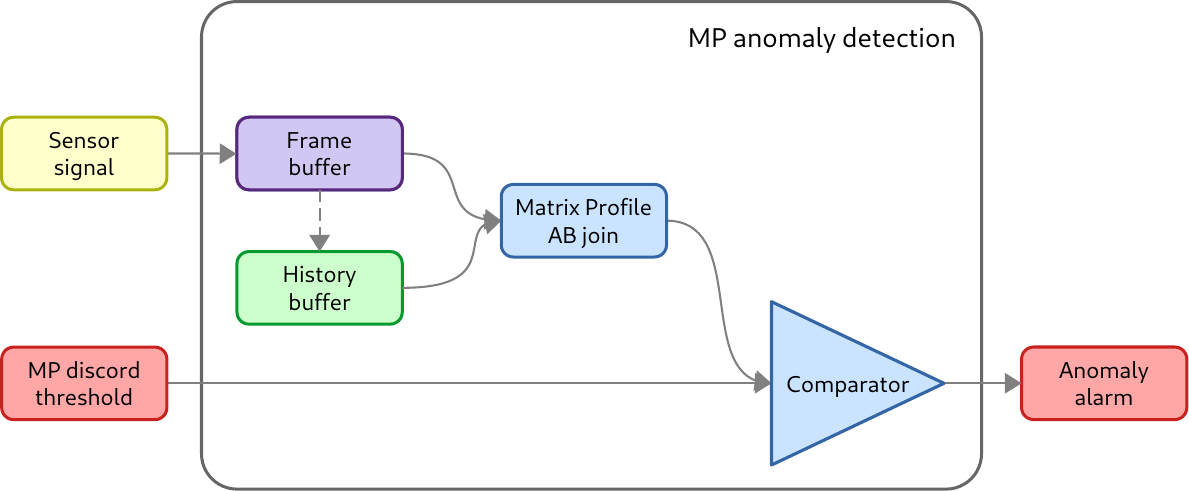}
   \caption{Architecture of the naïve implementation of a MP-based anomaly detection algorithm.}
   \label{fig:naive}
\end{figure}

We first explore the plausibility of using the MP for anomalous gait detection by implementing a naïve algorithm shown in Fig.~\ref{fig:naive}.
In it, signal samples are accumulated in a Frame buffer, which is updated at a specified hop length analogous to a sliding window.
As new samples are added to the Frame buffer, the oldest ones are transferred to a larger History buffer.
The contents of these two buffers overlap up to the specified exclusion zone for the MP algorithm (25\%).

For each update of the Frame and History buffers, the MP is calculated by using the Frame buffer to query the History buffer.
The value for the MP is then compared to a discord threshold and if larger the system activates an Anomaly alarm. 
The Frame buffer size, i.e. the subsequence length $m$, and the discord threshold are the two critical system parameters. 

\begin{figure}[t]
    \includegraphics[width=\columnwidth]{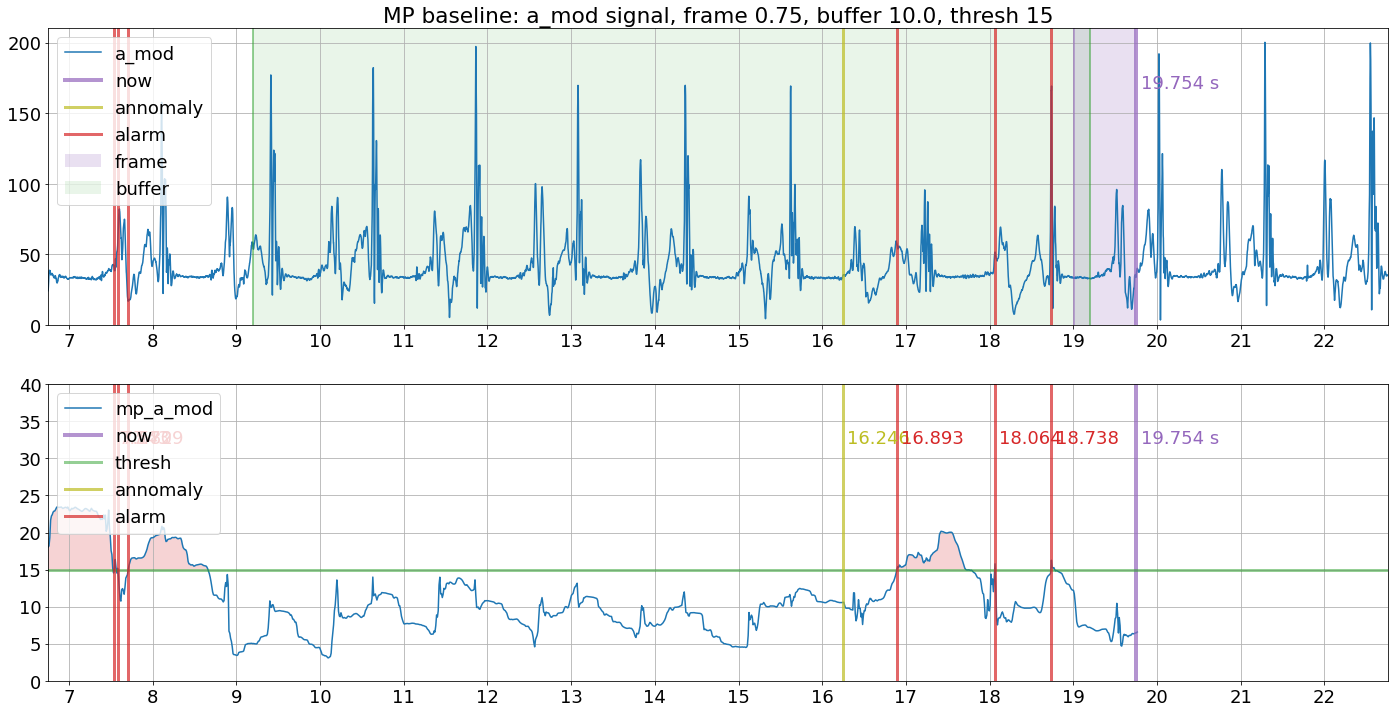}
    \caption{
        Visualization of the functioning of the naïve algorithm for a sample acceleration signal from the database (top plot) in which there are 7 normal steps followed by 2 anomalous steps and then 3 more normal steps. 
        The contents of the Frame and History buffers are highlighted in violet and green.
        The MP discord is calculated for each update of the Frame Buffer (bottom plot) and is compared to a threshold (green) raising an alarm if it goes above (red lines).
    }
    \label{fig:naive_signal}
\end{figure}

Fig.~\ref{fig:naive_signal} shows a qualitative inspection of the naïve algorithm for a sample acceleration signal.
We can see that the algorithm does indeed successfully detect the onset of anomalous steps raising an alarm, thus validating the approach.

\section{MP-based gait anomaly detection system}

\begin{figure}[t]
    \includegraphics[width=0.9\columnwidth]{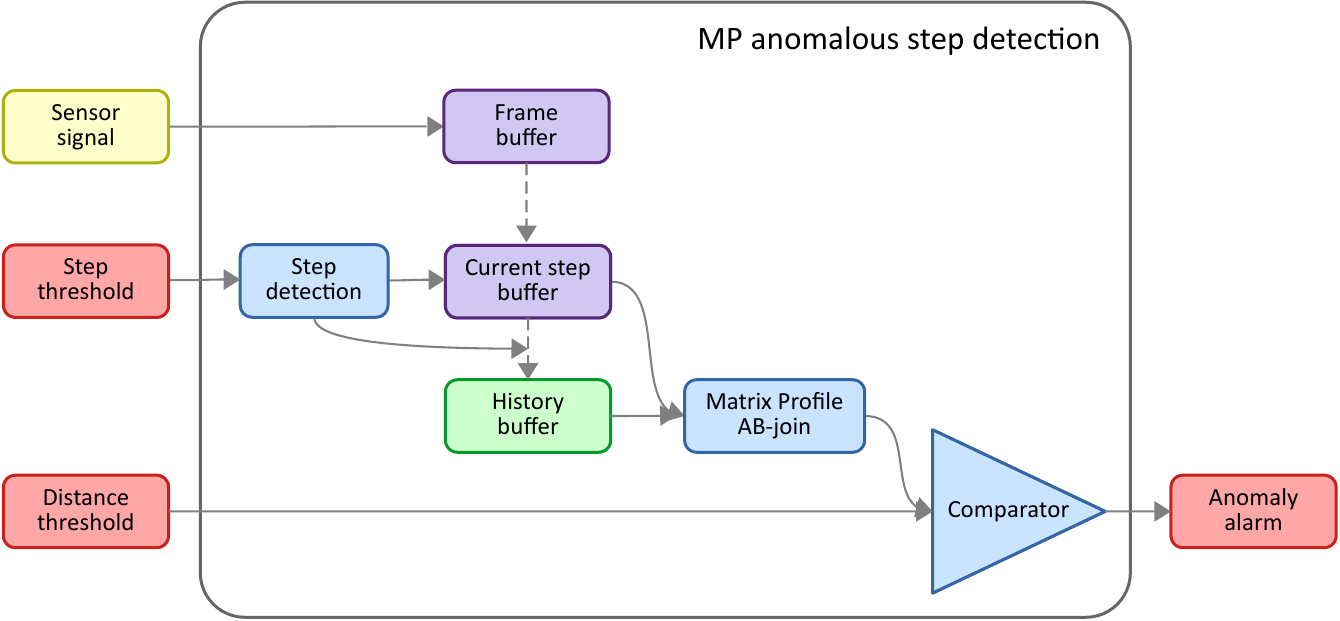}
   \caption{Architecture of the MP step based anomaly detection algorithm.}
   \label{fig:architecture}
\end{figure}

In the results from the naïve implementation, we can see that there is a problem at the start of the signal, where it generates false alarms.
This is because at this point in time the History Buffer does not contain any step signatures.
Based on our inspection, we designed an improved MP system architecture in which we integrate step detection, shown in Fig.~\ref{fig:architecture}. 
The input sensor signals are now forwarded from the Frame buffer and accumulated in a Current step buffer.
The step detection module analyses the contents of the Current step buffer on each update, and upon detecting the start of a new step it moves the contents of the Current step buffer to the History buffer.

The MP is calculated for each update of the Current step buffer, but only if the Step detection module has detected a step has started.
In this case, the subsequence length $m$ changes and is equal to the length of the signal stored in the Current step buffer. 

\subsection{Step detection algorithm}
\label{sec:step}

\begin{figure}
    \centering
    \includegraphics[width=\columnwidth]{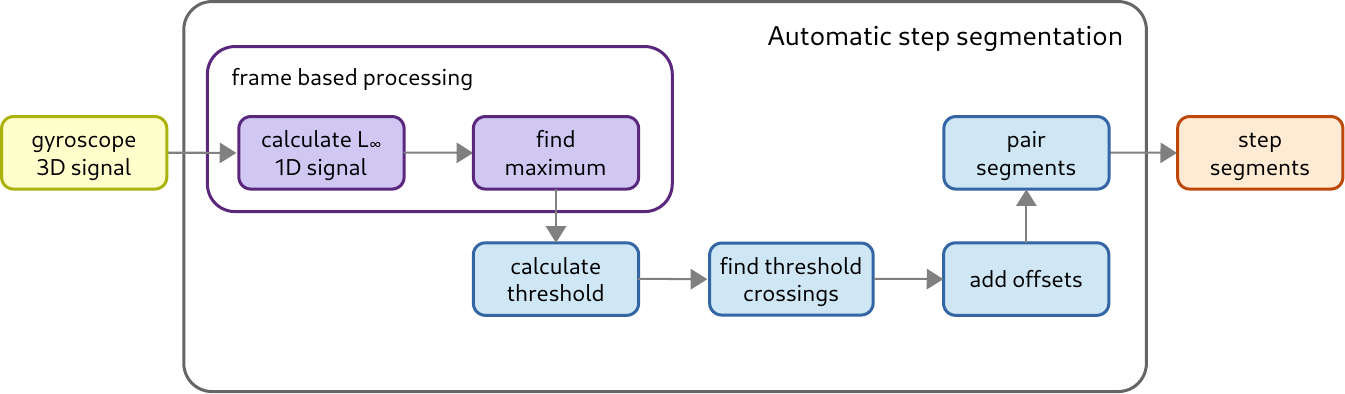}
    \vspace{5pt}

    \includegraphics[width=0.9\columnwidth]{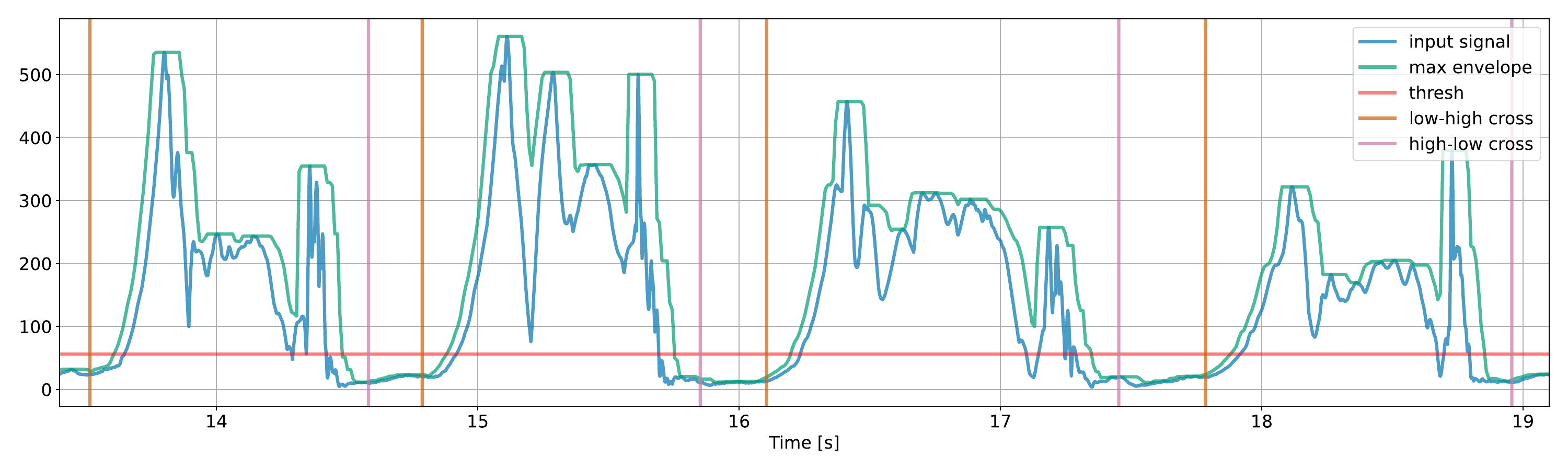}
    \caption{Block schematic of the step detection algorithm (top), and step detection results for a sample signal (bottom).}
    \label{fig:seg}
\end{figure}

The block schematic of our step detection algorithm is shown in the top plot of Fig.~\ref{fig:seg}.
It is based on an adaptive threshold that's used to detect crossings of the maximum amplitude envelope of the input signal.
Offsets are applied to the crossings to account for the step onset and release below the threshold.
The amplitude envelope is calculated with a wide 100~ms window that also acts as a low-pass filter. 
The threshold is adaptive and is recalculated with each update of the History buffer from the maximum value of the envelope signal stored in the History buffer.
In fact, setting the step segmentation threshold high initially, let’s the algorithm adapt only when actual steps are buffered in the History buffer. 
The results from using it on the sample signal are shown in the bottom plot.
In a subset of experiments we determined that the $L_\infty$ 1D projection of the gyroscope signal gives the best step segmentation results.

\section{Experiments}

We conducted a set of experiments to optimize and evaluate the proposed system.



{\bf Plausibility.}
As with the naïve implementation, we qualitatively evaluated our MP-based gait anomaly detection system with sample signals from our database.

{\bf Sensor signal.}
We analyzed the performance of the MP algorithm when the three different axes of the gyroscope and accelerometer signals are used, and their $L_1$, $L_2$ and $L_\infty$ norms.

%
%
{\bf External signals as reference.} 
We evaluate the possibility of using preset normal steps from external sources as reference in the History buffer. 
This has the potential to ease deployment, but comes at the cost of curbing adaptation.
Here, we make two subexperiments: 1) extracting the reference from the diplegic/hemiplegic signals, and 2) using a mix of segments from all anomalies, 10~s each. For a fair comparison we also use increased lengths of the History buffer.

{\bf Neural Network baseline.}
To evaluate the comparative performance of our proposed algorithm we design, train and optimize a Neural Network baseline system based on recurrent LSTM (long short-term memory) layers. 
The optimized architecture of the model comprises two layers of bidirectional LSTMs with a size of 256, followed by a 3-hidden layer feedforward network, sizes 256, 128, and 64, and a final output neuron with a linear activation function.
All layers in the network were followed by batch normalization and a dropout of 0.2.
The network was fed 2~s of the Sensor signal.

While for the other experiments we use a smaller subset of the data for efficiency, here we use a larger proportion to get a better estimate for in-the-wild performance.


\begin{figure}[t]
    \includegraphics[width=\columnwidth]{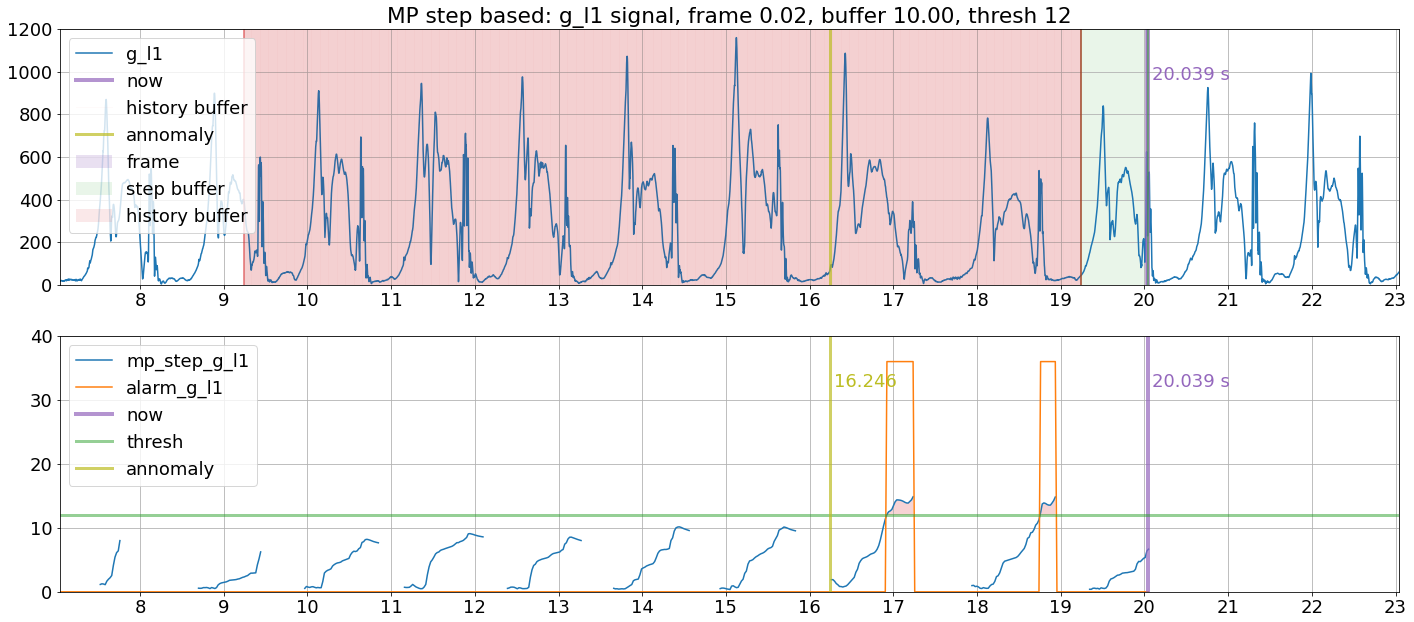}
    \caption{Visualization of the functioning of the MP-based baseline system for gait anomaly detection.}
    \label{fig:full-history}
\end{figure}

{\bf Evaluation metrics.}
To evaluate the performance of our proposed system we employed metrics commonly used in binary classification tasks including the F1 score, ROC (Receiver Operating Characteristic) and earliness, i.e. the average latency in seconds needed for the system to raise an alarm upon the onset of an anomalous step. 

\section{Results}

{\bf Plausibility.}
Fig.~\ref{fig:full-history} shows the contents of the Current step and History buffers as well as the calculated MP and detected alarms for the $L_1$ norm of a sample gyroscope signal.
We can see that indeed the MP algorithm is capable of detecting anomalous steps, and also and deals efficiently with the start of the signal.

\begin{figure}
    \centering
    \includegraphics[width=0.8\columnwidth]{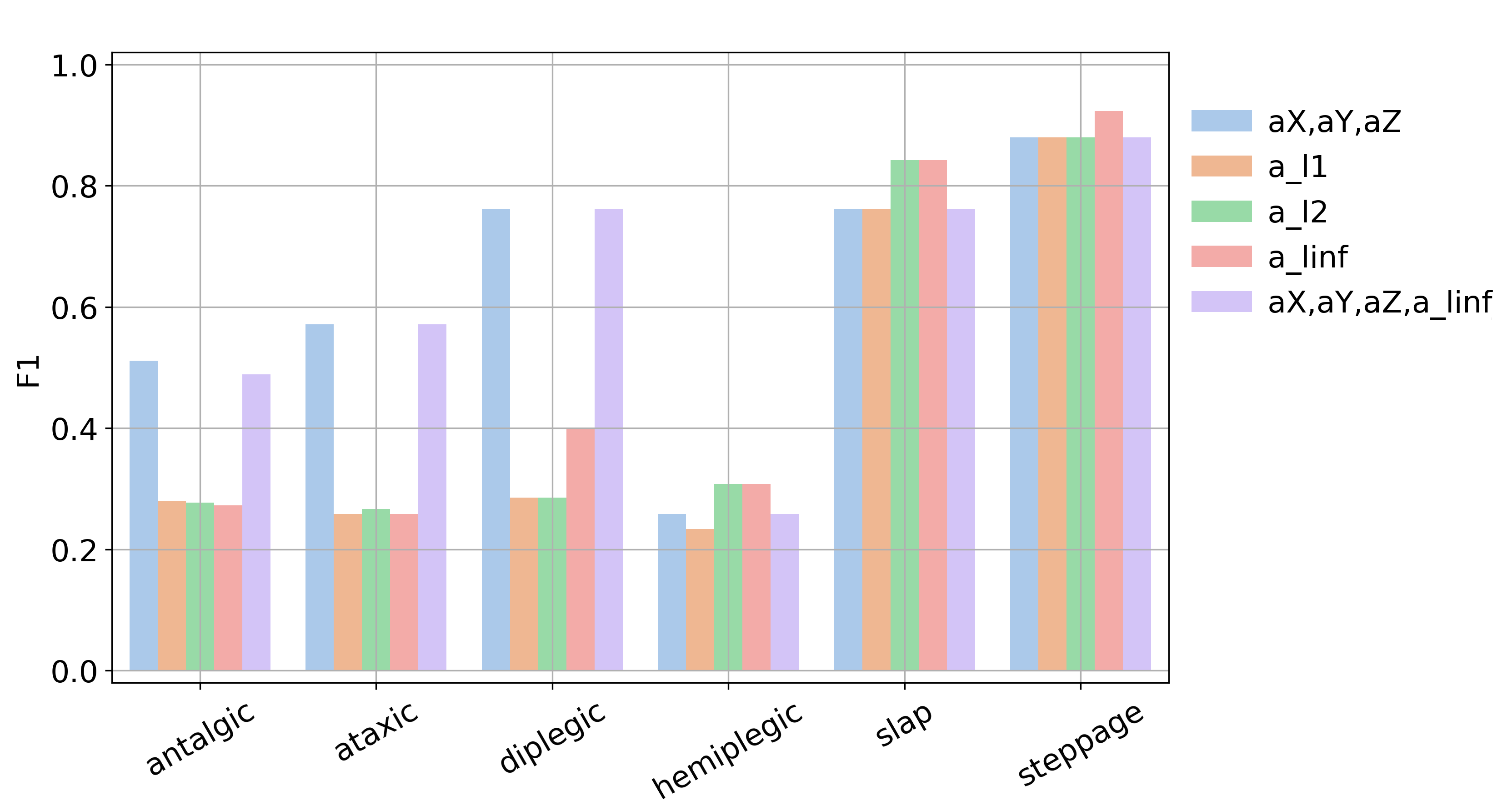}
    \includegraphics[width=0.8\columnwidth]{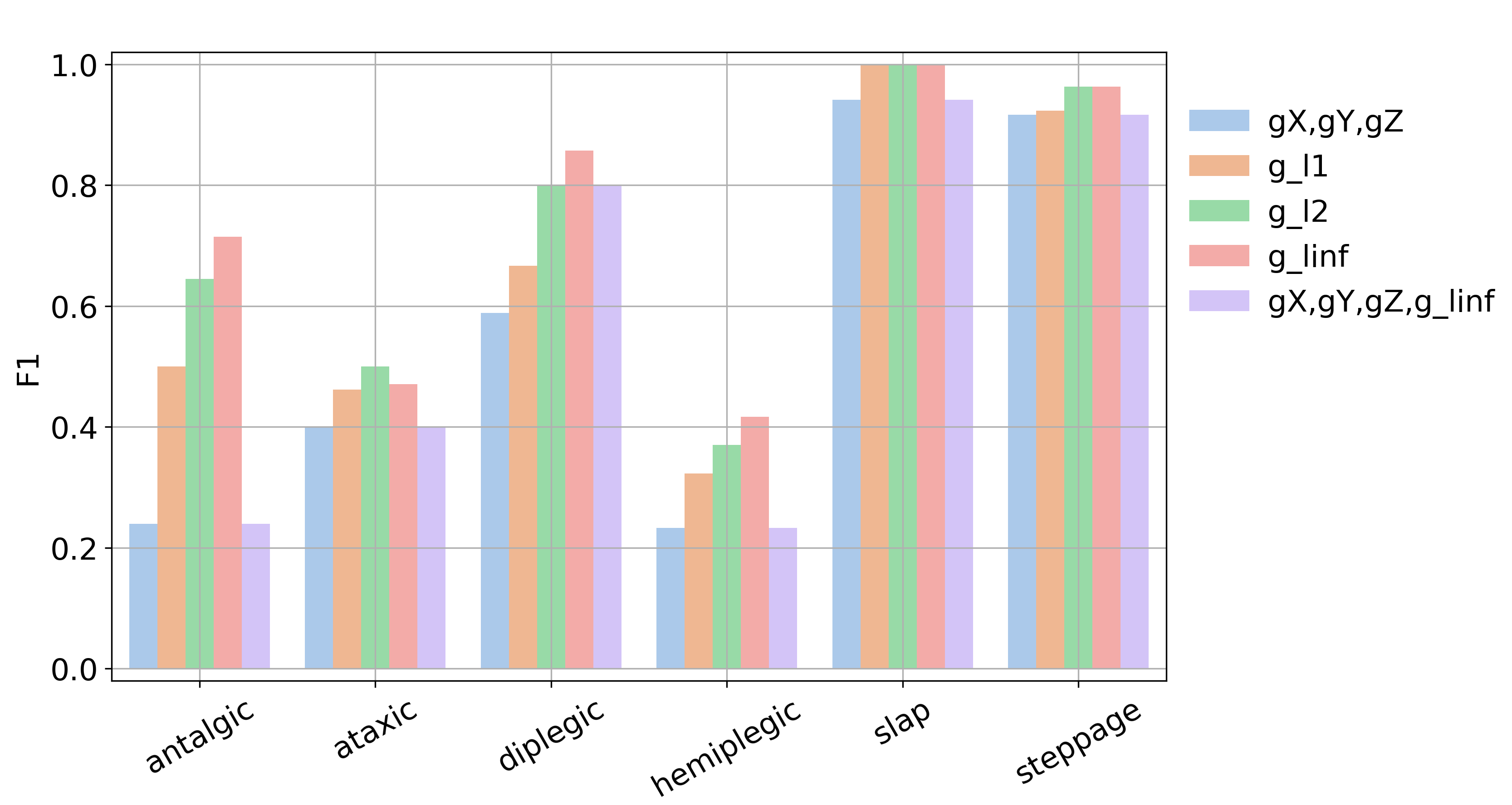}
   \includegraphics[width=0.7\columnwidth]{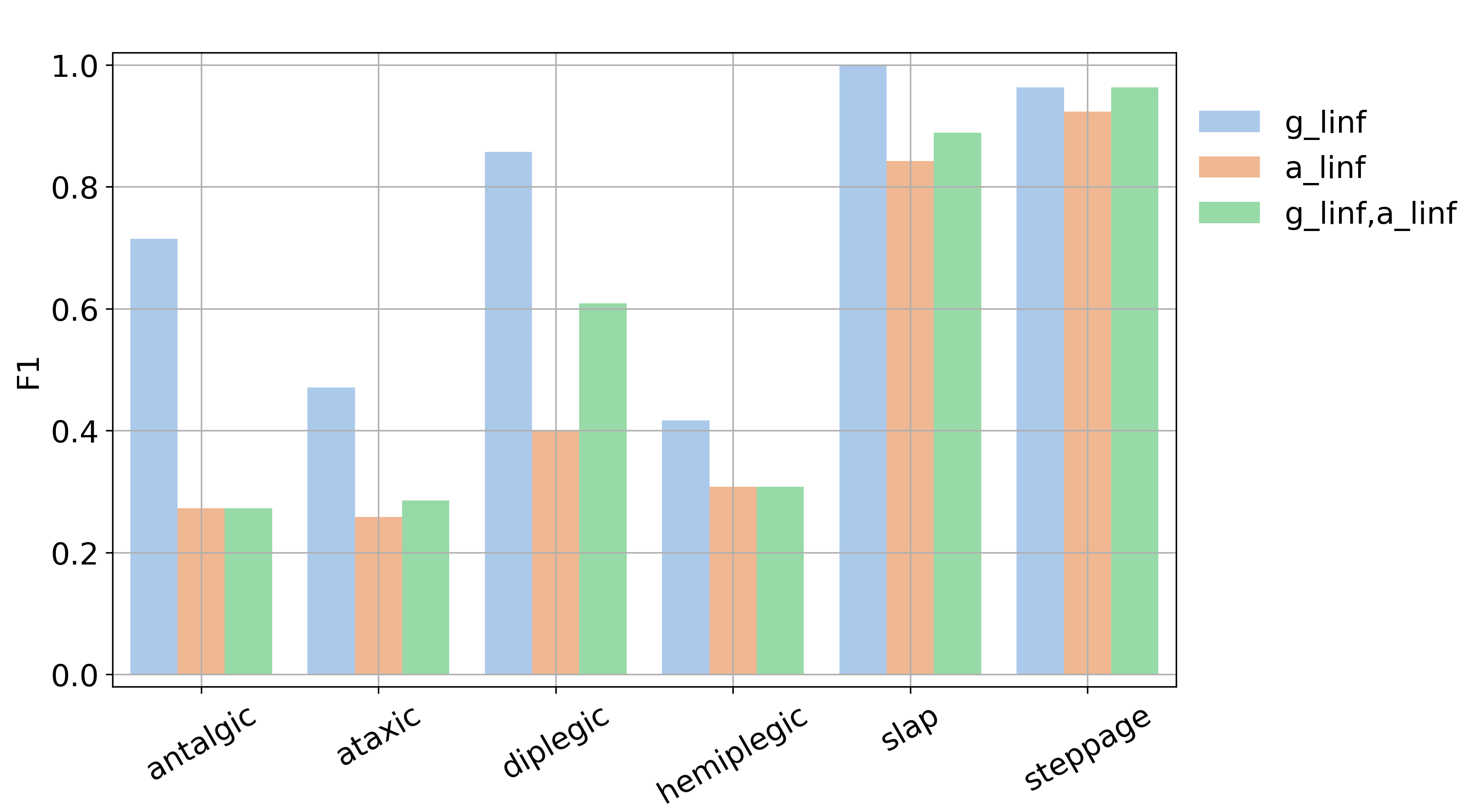}
    \caption{Best case $F1$ - score for different axes and norms from the accelerometer (top), gyroscope (middle), and both (bottom) signals.}
    \label{fig:best_f1_gyro_axes}
\end{figure}

{\bf Sensor signal.}
The F1 results for the accelerometer and gyroscope signals, across all anomalies, are shown in Fig.~\ref{fig:best_f1_gyro_axes}.
The relative F1 really varies across the anomalies for the accelerometer signal.
For the gyroscope signal they are more consistent, with the $L_\infty$ norm performing better, while using or adding multiple axes, degrades performance.
In the bottom, we can see that the gyroscope $L_\infty$ norm outperforms the accelerometer $L_\infty$ norm, as well as when both signals are used.

%
%

\begin{figure}
    \includegraphics[width=0.85\columnwidth]{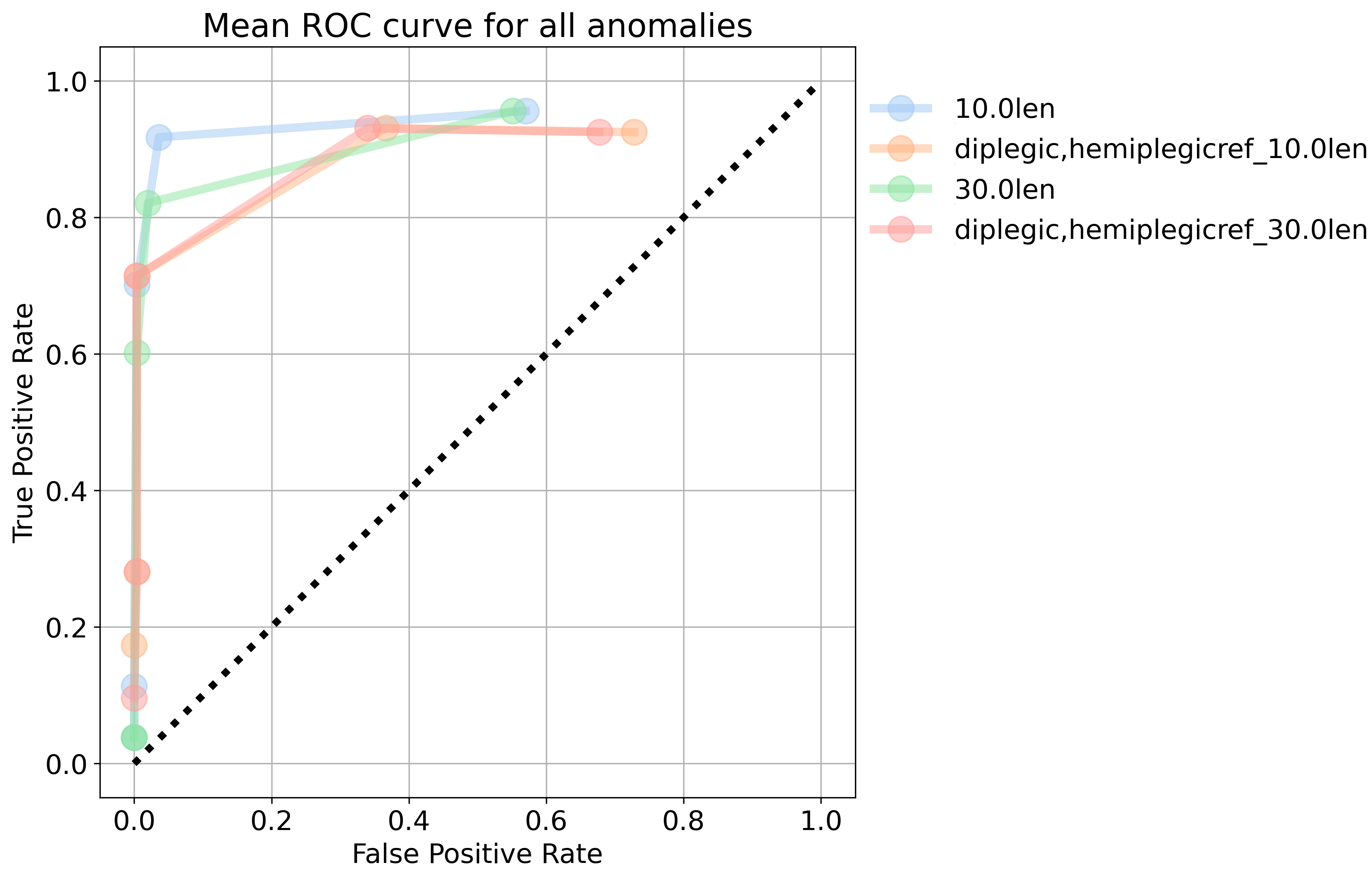}
    \vspace{8pt}\null

    \includegraphics[width=0.85\columnwidth]{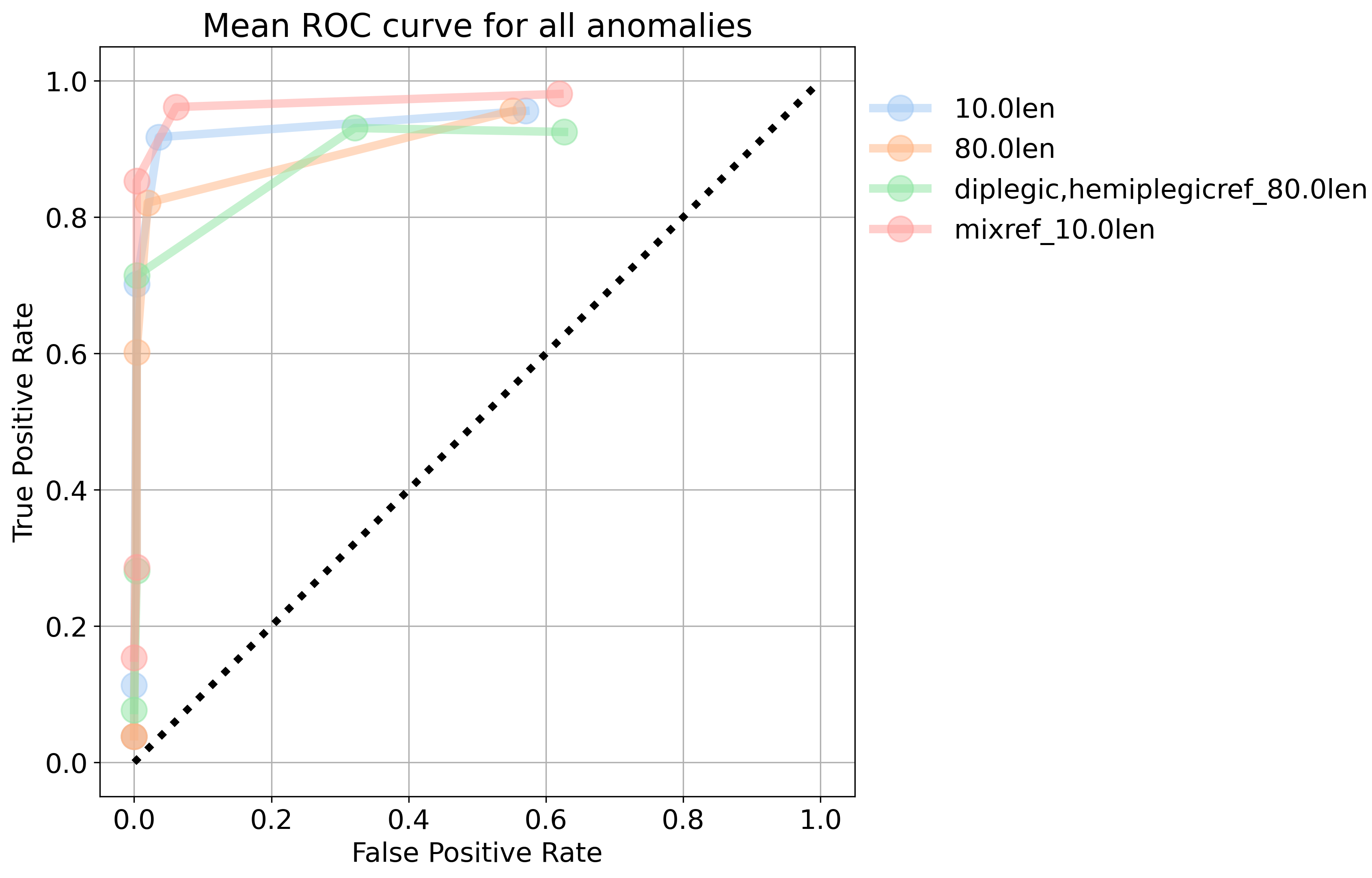}
    \caption{Mean ROC for the MP-based system for different lengths of the History buffer and different signals used as reference.}
    \label{fig:roc_10vs30}
\end{figure}

\begin{figure}[t]
    \centering
    \includegraphics[width=0.75\columnwidth]{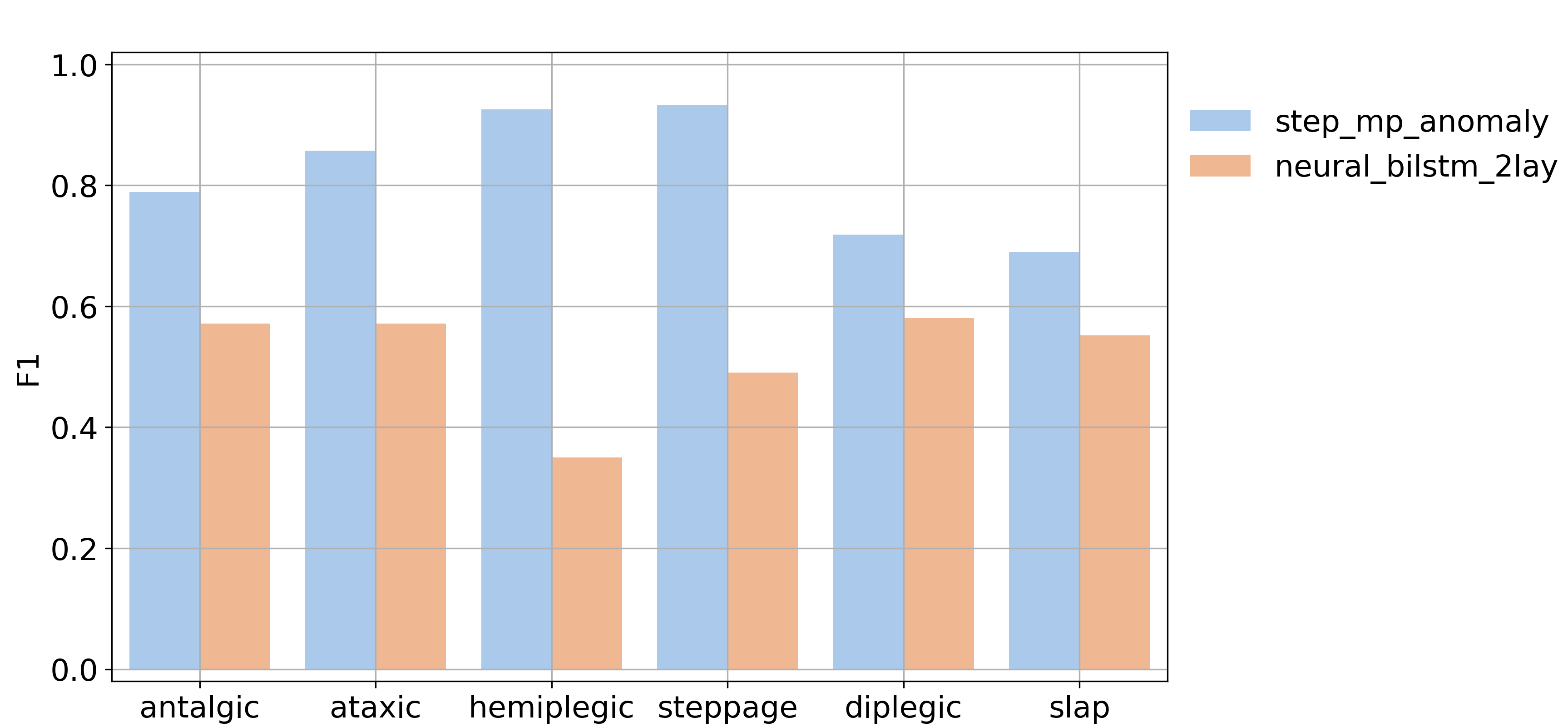}
    \includegraphics[width=0.75\columnwidth]{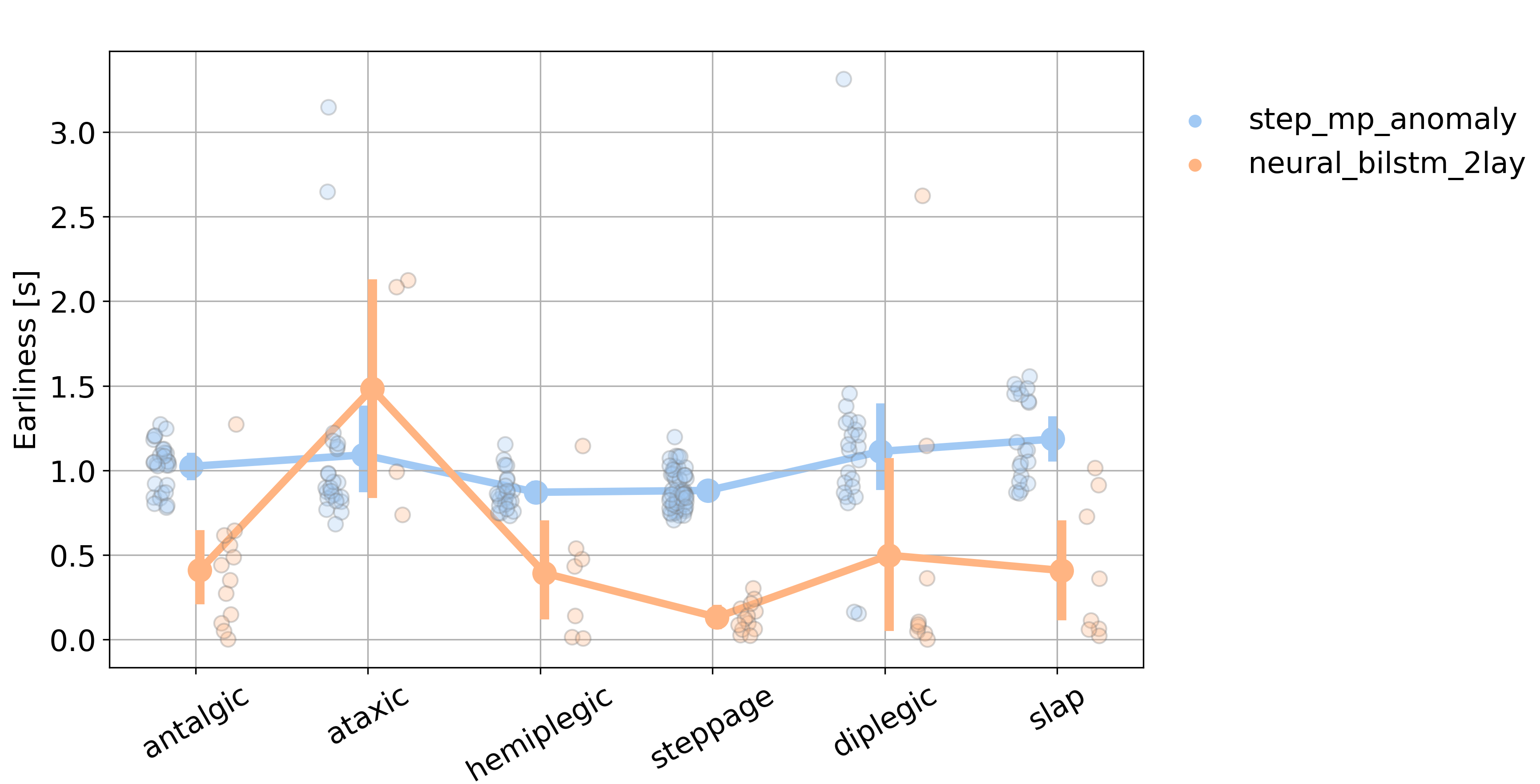}
    \caption{Mean F1 (top) and earliness (bottom) for the MP-based system compared to the Neural Network baseline.}
    \label{fig:f1neural}
\end{figure}

{\bf External signal reference.} 
The results from using different lengths of the History buffer and different signals used as reference are shown in Fig.~\ref{fig:roc_10vs30}.
Comparing the mean ROC curves, we can see that on average, there is benefit of using a mixed signal reference.
Closer inspection however, omitted here for brevity, shows that the results vary by anomaly.


{\bf Neural Network baseline. }
The F1 results comparing the Neural Network baseline to the proposed model are shown in the top of Fig.~\ref{fig:f1neural}.
It can be seen that the MP-based algorithm outperforms the Neural Network baseline by a wide margin.
The Neural Network does provide faster reaction times than the MP-based system as can be seen in the bottom plot. 
We also measured the real-time factor of the two algorithms and found that it is 10$\times$ higher for the Neural Network baseline. 
This might point towards possible deployment issues on edge devices.

\section{Conclusion} \label{Conclusion}

We propose a gait anomaly detection system based on the Matrix Profile algorithm.
The system relies on lean digital signal processing to adapt to an individual's gait pattern and to successfully detect outliers with low latency.
The system obtains high F1 scores across anomalies, outperforming a more complex Neural Network baseline.
Its low complexity makes the MP based gait anomaly detection system a good candidate for edge deployment.

\bibliographystyle{IEEEtran}
\bibliography{IEEEabrv,refs}

\end{document}